# Biography of the French astronomer Henri Camichel

Emmanuel Davoust, *Observatoire Midi-Pyrénées, F-Toulouse*

*Abstract*. Henri Camichel was an astronomer at Pic du Midi Observatory, where he contributed to the study of planets of the solar system and their satellites with Audouin Dollfus and his team. In 1961, with Charles Boyer, he found that the upper atmosphere of Venus had a counter-clockwise rotation of four days, which was later confirmed by space probes, as were the team's accurate measurements of the diameters of planets. He was also an instrumentalist, and contributed to the maintenance and development of the telescopes notably the 2-meter telescope and focal instruments at Pic du Midi Observatory.

Henri Camichel was born on December 23, 1907 in Roquefort near Agen, in south-western France, where his father was a doctor in the French army. He became interested in astronomy early on, and acquired an 8cm refractor in the 1920s. He studied astronomy under Emile Paloque at Toulouse University and under Jean Cabannes in Montpellier University, where he obtained a Bachelor of Science degree. In the fall of 1930, he became an unpaid intern at Paris Observatory where he contributed to observations of the minor planet Eros at the meridian circle. Once familiar with that instrument, he joined the Institute of Astrophysics of Paris where he researched a possible variation of the constant of aberration with spectral type. During his military service in 1932-1933, he worked on locating artillery batteries by sound. In the spring on 1933, he joined Meudon Observatory, again as an unpaid intern, where he conducted solar observations for Lucien d'Azambuja. He also wanted to observe nights, and worked with Eugène Antoniadi at the 83cm refractor, which is perhaps how he developed an interest in Mars and planetary astronomy. In December 1934, at the urge of Charles Bertaud, he observed Nova Herculis with a spectrometer at the 83cm refractor (Camichel 1996).

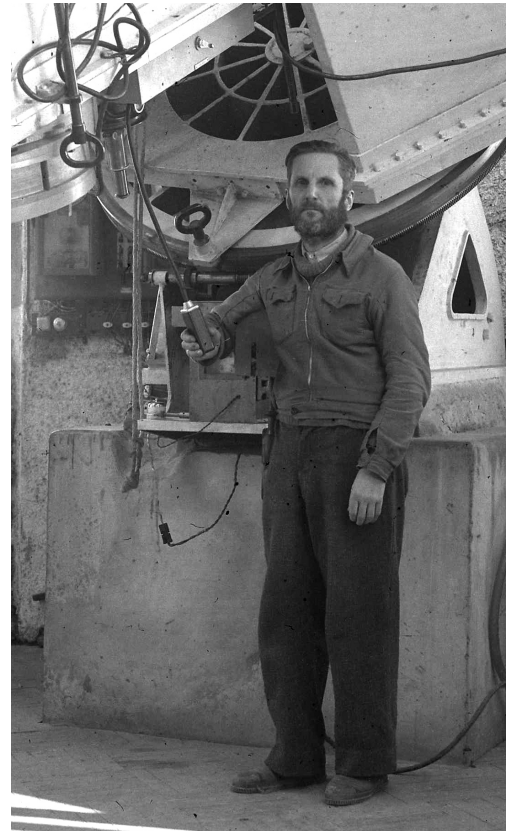

*Henri Camichel in front of the refracto-reflector at Pic du Midi Observatory in 1956*

Camichel's colleague Bernard Lyot invited him to observe the solar corona at Pic du Midi Observatory in the summer of 1936. Another colleague, Fernand Baldet, shared with Camichel his experience using the Pic du Midi telescope in 1909 and 1910. In November of that year1936, Camichel joined the Observatory staff as a physicist in charge of meteorological observations. Since these tasks were not very demanding, he started observing at the 50cm Benjamin Baillaud telescope, which until then had never been used in winter. He was mobilized at the outbreak of WWII as a lieutenant in a group locating enemy artillery batteries, but was stationed at Pic du Midi in November of 1939 to monitor the propagation of metric waves at high altitude. He resumed his observations of Mars after the armistice.

During the war years, Lyot observed the solar corona at Pic du Midi every summer. At Lyot's initiative and with his help, Camichel modified the telescope which had a mediocre mirror.

In 1941 he mounted a 38cm objective borrowed from Toulouse Observatory on the tube of the telescope. In 1943, he transformed the 50cm reflector and its attached 23cm viewfinder into a refracto-reflector, using the 60cm objective of the Paris large Coudé telescope and a 50cm plane mirror. The telescope was then used in that configuration for over forty years, for solar and planetary research. Camichel took advantage of the excellent seeing to obtain photographic sequences of Mars, Uranus, Saturn, Jupiter and Mercury, planets for which the surface and period of rotation were poorly known.  Camichel used the method of plate stacking to improve the resolution of the images. Lyot also had an ambitious project of a domeless 150cm telescope for the Pic du Midi Observatory in which Camichel participated, a project which finally aborted for lack of funds.

Right after the war, Audouin Dollfus from Meudon Observatory formed a team of observers of planets at Pic du Midi, including Camichel "the instrument man", as he called him, Greek-French astronomer Jean Focas and Italian astronomer Glauco de Mottoni (Dollfus 2003). This team produced valuable data on the planets until the advent of space probes (e.g.  Camichel and Dollfus 1968). Camichel defended a thesis in Paris on the determination of the pole, diameter and areographic coordinates of Mars in 1952, using his collection of photographic plates taken over a decade at Pic du Midi (Camichel 1954). He participated in a project led by Jean Rösch, the director of the Observatory, to measure the diameter of Mercury by observing its transit over the sun in 1960 and 1970.  In 1957 he started a collaboration with an amateur astronomer,  Charles Boyer, who observed with a 20cm telescope in Abidjan Ivory Coast, to observe Venus. In 1961, using photos in the ultraviolet, they determined the period of rotation of a Y-shaped feature on the surface of Venus, which was rotating counterclockwise in four days (Boyer and Camichel 1961, 1967). This discovery was confirmed by Mariner 10, which made a flyby in 1974.

Camichel became heavily involved in the project for the construction of a 2-meter telescope at Pic du Midi Observatory, which started in 1968. He retired in 1977, one year before the first light of the telescope. He had planned to take photos of Pluto, and later speculated that, if the telescope had not suffered several delays, he might have discovered, thanks to the high resolution of the site and the collecting power of the instrument, that Pluto had a companion.

After a first marriage in 1943 which ended in divorce in 1947, Henri Camichel married Claire Galibert in 1966. The couple remained childless. In retirement Camichel became interested in French history  and reflected on his career in the exceptional site of Pic du Midi, while enjoying the large garden of his house next to Toulouse Observatory. He passed away in Toulouse on January 12, 2003

Camichel published 34, mostly refereed, papers. He was awarded the "Prix Lalande" of the French Academy of Science in 1942 for his work in physical astronomy. A crater 65.3km in diameter on Mars was named after him in 2012 by the International Astronomical Union, in recognition of his contribution to the study of that planet.

**Selected bibliography**

Camichel, H. 1954. Détermination photographique du pole de Mars, de son diamètre et des coordonnées aréographiques.  *Bulletin Astronomique* 18 : 83-191

Boyer, Ch., Camichel, H. 1961. Observations photographiques de la planète Vénus, A*nnales d'Astrophysique* 24 : 531-535

Boyer, Ch., Camichel, H. 1967. Détermination de la vitesse de rotation des taches de Venus. *Comptes Rendus de l'Académie des Sciences, Série B Sciences Physiques* 264 : 990- 992,